\documentclass[fleqn,letterpaper,11pt]{article}
\usepackage{graphicx,physics,xcolor,bm,bbm, amsmath,amsfonts,amsbsy,amssymb,fullpage,longtable}
\usepackage[numbers,sort&compress]{natbib}
\usepackage[title]{appendix}
\usepackage[hidelinks,colorlinks=true,linkcolor=blue,citecolor=blue,urlcolor=blue]{hyperref}
\usepackage{orcidlink}  

\newcommand{\eq}[2]{\begin{gather} #2  \label{#1} \end{gather}}
\newcommand{\nn}{\nonumber\\} \newcommand{\Po}{\text{Po}} \newcommand{\Nu}{\text{Nu}} \newcommand{\Br}{\text{Br}}

\begin{document}

\title{\bf Wall roughness and viscous dissipation effects in microchannel heat sinks with semicircular cross-section}

\author{\bf A. Barletta$^{*}$\orcidlink{0000-0002-6994-5585}, M. Celli\,\orcidlink{0000-0002-2726-4175}, P. V. Brandão\,\orcidlink{0000-0002-8445-3802}}

\date{\small
{\em Department of Industrial Engineering, Alma Mater Studiorum Università di Bologna,\\ Viale Risorgimento 2, 40136 Bologna, Italy}\\[0.4cm]
$^*${corresponding author:\ {\tt antonio.barletta@unibo.it}}
}
\maketitle

\begin{abstract}\noindent
A statistical analysis of the wall roughness effect is carried out to determine the impact of the shape uncertainty on the Poiseuille number and Nusselt number of laminar forced convection. The focus is on the fully developed regime in a semicircular microchannel where the heat transfer occurs from the diametrical plane boundary, modelled as a perfectly smooth surface. On the other hand, the curved semicircular boundary is devised as rough and with a negligible wall heat flux. Three types of thermal boundary conditions are implemented: the {\sf T} condition, the {\sf H1} condition and the {\sf H2} condition. The {\sf T} condition serves to model a case where the fluid temperature does not undergo any change in the streamwise direction, while the {\sf H1} and {\sf H2} conditions are employed to describe a net heating of the fluid. A statistical sample of several different rough microchannels is used to detect the actual effects of roughness on the Poiseuille number and on the Nusselt number, through the evaluation of their average values and standard deviations. The governing local momentum and energy balance equations are solved numerically by a finite element method taking into account the viscous dissipation contribution to the local energy balance. \\[12pt]
\end{abstract}

\subsection*{Nomenclature}
\begin{longtable}[l]{ll}
$a$ & constant temperature gradient\\
$\Br$ & Brinkman number\\
$c$ & specific heat\\
$k$ & thermal conductivity\\
$n$ & integer\\
$\vu n$ & unit normal vector\\
$N$ & number of nodes along the polygonal\\
$\Nu$ & Nusselt number\\
$p$ & pressure\\
$\cal P$ & cross-sectional perimeter\\
${\cal P}_1, {\cal P}_2$ & nominally semicircular boundary, diametrical boundary\\
$\Po$ & Poiseuille number\\
$q_w$ & average wall heat flux\\
$r$ & radial coordinate\\
$r_h$ & nominal hydraulic radius of the smooth semicircular duct\\
$r_0$ & radius of the semicircle\\
$\cal S$ & cross-sectional area\\
$T$ & temperature\\
$T_b$ & bulk temperature\\
$T_w$ & average wall temperature\\
$u$ & velocity\\
$u_0$ & reference velocity\\
$u_m$ & mean flow velocity\\
$x,y,z$ & Cartesian coordinates\\
$x_n, y_n$ & coordinates of the polygonal points\\[6pt]
{\em Greek symbols} \\
$\alpha$ & thermal diffusivity\\
$\gamma$ & roughness dimensionless parameter\\
$\delta r_n, \theta_n$ & radius uncertainty, angular position of the $n$th polygonal point\\
$\mu$ & dynamic viscosity\\
$\nu$ & kinematic viscosity\\
$\sigma$ & dimensionless streamwise temperature gradient\\[6pt]
 {\em Subscripts, superscripts}\hspace{-1.5cm} \\
$^*$ & dimensionless quantity\\
{\sf T}, {\sf H1}, {\sf H2} & thermal boundary conditions\\
\end{longtable}
\setcounter{table}{0}

\section{Introduction}
The use of microchannels is fundamental in modern thermal management systems, especially in applications requiring compact and efficient heat removal such as microelectronics, microreactors, and biomedical devices \citep{khan2011review, kandlikar2012history, siddiqui2017efficient, gilmore2018microchannel, deng2021review, harris2022overview, heidarshenas2024enhancing}. Microchannels are typically characterized by hydraulic diameters usually less than $100 \mu$m. When a fluid is actively driven through these internal passages with a pressure gradient supplied by external pumping, the resulting convective heat transfer takes place in a forced convection regime. Due to the high area-to-volume ratio of microchannels, they enable efficient heat exchange even with relatively low flow rates, making them ideal for miniaturized heat sinks \citep{dash2022role, yu2024comprehensive, DONG2024125001}.

In microchannels, the small scale of the flow passages introduces several complex physical phenomena that are either negligible or less significant in conventional-sized channels. Among the most critical of these are viscous dissipation and wall roughness, both of which can significantly affect the heat transfer and fluid flow behaviour \citep{ BXu2003, KOO20043159, CROCE2004601, croce2005numerical, bruus2007theoretical, croce2007three, van2009effect, turgay2009effect, Nonino2010, pelevic2016heat, mukherjee2017effects, jing2017joule, lu2020effects}.

Viscous dissipation is the irreversible conversion of mechanical energy into heat due to viscous friction within the fluid. In microchannels, where the flow velocities may be large relative to the small cross-sectional area, viscous heating effects are amplified. As a result, the fluid experiences a significant temperature rise due to internal friction. The presence of viscous dissipation adds a source term to the energy equation which can lead to increased fluid temperatures, altering the temperature gradient between the wall and the fluid, and thus impacting the overall heat transfer rate \citep{BARLETTA199615, bruus2007theoretical}.

No real microchannel has perfectly smooth walls. Surface roughness due to manufacturing tolerances, etching artifacts or deliberate texturing alters both hydrodynamic and thermal behaviour. At the microscale, a roughness element of just a few microns can occupy a significant fraction of the channel’s hydraulic diameter. Thus, unlike in conventional systems, where wall roughness may be treated as a minor perturbation, in microchannels, even small-scale surface irregularities can have a disproportionately large influence on the flow and heat transfer characteristics. Manufacturing techniques such as micromachining, etching, or 3D printing can introduce surface imperfections that disturb the fully-developed laminar flow profile typically expected in smooth microchannels. Roughness may cause an increase in the friction factor, leading to a higher pressure drop for a given flow rate. From a heat transfer standpoint, roughness may enhance heat transfer due to an increased fluid-to-solid exchange area or it can even induce a decrease in the heat transfer rate due to the flow velocity reduction in the near-wall region \citep{wang2007influence, croce2007three, turgay2009effect, pelevic2016heat}. 

Forced convection in microchannels is highly sensitive to effects that are often negligible in macroscale systems. Viscous dissipation introduces additional heating, affecting temperature distributions and potentially reducing thermal control accuracy. Wall roughness modifies flow characteristics and can either enhance or contrast heat transfer depending on the roughness geometry and flow conditions. Accurate modelling of microchannel systems must consider these phenomena to ensure optimal design and operation in high-performance thermal applications.

Viscous dissipation and wall roughness interact: rougher surfaces elevate shear rates locally, further amplifying viscous heating \citep{barletta2025shape, sphaier2025laminar}. Consequently, optimization of microchannel heat sinks involves a careful balancing act. Computational fluid dynamics (CFD) simulations and micro-PIV (particle image velocimetry) experiments help map out the parameter space of Reynolds number, Brinkman number and roughness geometry. Material selection, fabrication method ({\em e.g.}, plasma-based DRIE, micromilling, 3D printing) and coolant properties are chosen to minimize detrimental effects while maximizing heat transfer density. In many applications, a moderate, controlled roughness pattern, such as staggered micro-rib arrays, provides enough mixing to enhance thermal performance without incurring prohibitive pressure drops. At the same time, low-viscosity, high-thermal-conductivity fluids (e.g., water-glycol mixtures or nanofluids) are preferred to control the effects of viscous dissipation.

\begin{figure}
\centering
\includegraphics[width=0.7\textwidth]{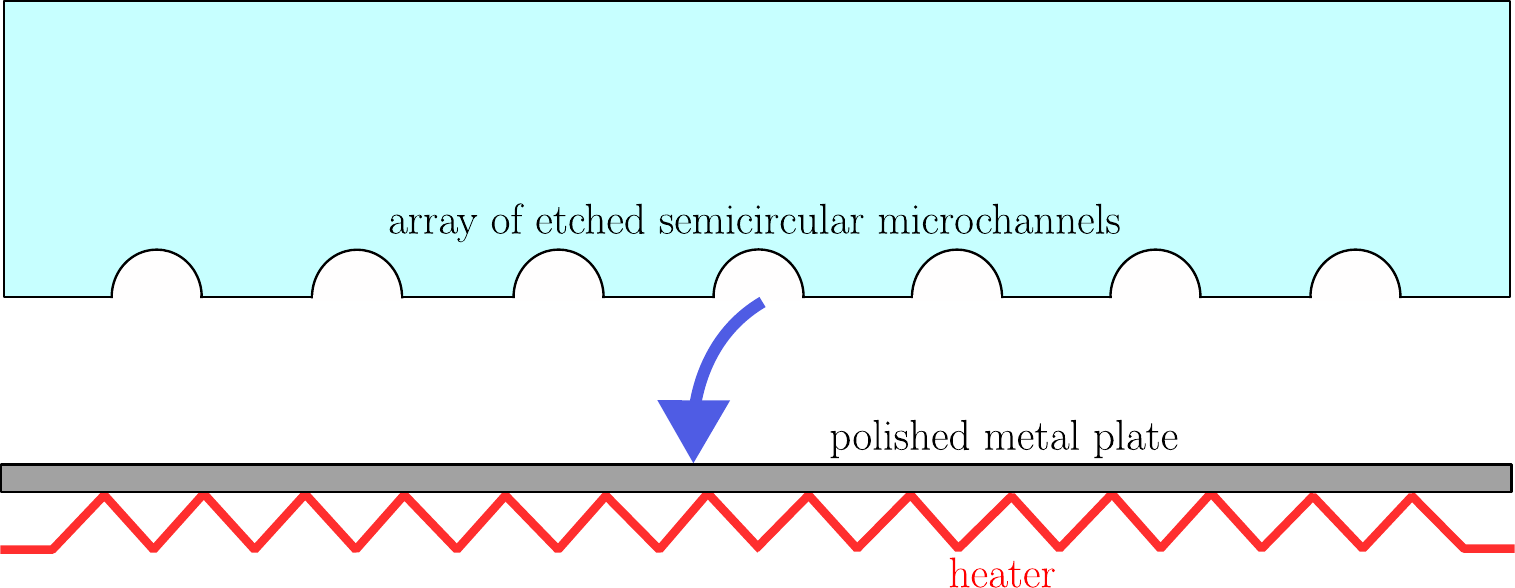}
\caption{\label{fig1}Cross-sectional view of an array of semicircular microchannels employed to cool a heated plate }
\end{figure}

The aim of this paper is the analysis of the mechanical and thermal characteristics of the forced convection flow across an array of semicircular microchannels employed to cool a heated polished metal plate. Such a simplified heat sink design is meant to better understand the actual heat transfer process involved in the cooling of an electronic circuit board. The basic scheme is reported in Fig.~\ref{fig1}. We assume an array of parallel microchannels etched on the surface of a solid layer. By sealing the layer on the heated metal plate, we devise a heat sink system where the thermal energy from the metal plate is convected away through the microchannels. We also assume that the thermal characteristics of the heat sink can be consistently inferred from the behaviour of a single microchannel, meaning that the thermal interaction between neighbouring microchannels is considered as negligible. This study develops the ideas recently discussed in the case of circular microducts by \citet{barletta2025shape} and \citet{sphaier2025laminar}. In particular, the approach where the roughness effect is examined on a statistical basis by simulating numerically the forced convection flow in randomly generated cross-sectional geometries is here applied to the case of semicircular microchannels.
In Section~\ref{mamo}, with reference to a single microchannel, the physical assumptions underlying our analysis and the mathematical model adopted are described in details. In Section~\ref{nondim}, the nondimensional analysis of the model developed is carried showing up the main dimensionless parameters governing the flow and the heat transfer process, providing also a description of the finite-element solver procedure employed for the numerical analysis of the mathematical model. Section~\ref{smoothscd} is devoted to the analysis of the smooth semicircular duct, namely the special case where the surface roughness is considered as negligible. This special case is also employed as a test for the assessment of the numerical solver accuracy. The effects of the wall roughness and its interplay with the viscous dissipation effect are drawn in Section~\ref{diswalrou}.

\section{Mathematical model}\label{mamo}
\begin{figure}
\centering
\includegraphics[width=0.8\textwidth]{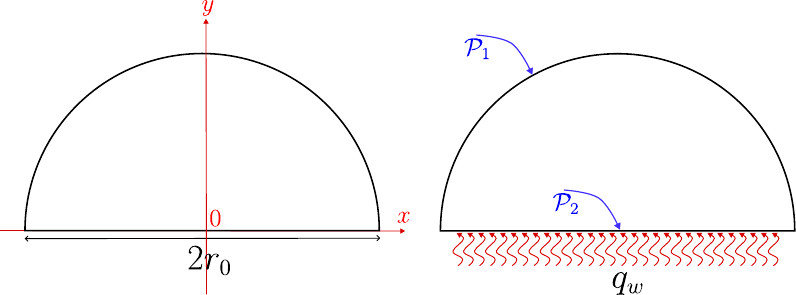}
\caption{\label{fig2}Sketch of the microchannel nominal cross-section}
\end{figure}

Let us consider a microchannel with a semicircular cross-section heated through its diametrical boundary side (see Fig.~\ref{fig2}). On the other hand, the semicircular boundary side is assumed to be adiabatic, meaning that the main heat transfer mechanism is the forced convection taking place in the axial $z$ direction. This simplifying assumption is realistic especially when the microchannel is obtained by etching a solid layer with poor thermal conductivity and/or with a thickness much larger than the microchannel diameter. The diametrical side is the contact surface with a heated metal plate. Depending on the fabrication characteristics of microchannels, we assume that the semicircular side is endowed with a roughness significantly larger than that of the heated metal plate, so that the latter will be modelled as perfectly smooth throughout this study. The nominal radius of the semicircular region is $r_0$. The semicircular region lies in the $xy$ plane, while the $z$ axis is parallel to the streamwise direction.

\begin{figure}
\centering
\includegraphics[width=0.4\textwidth]{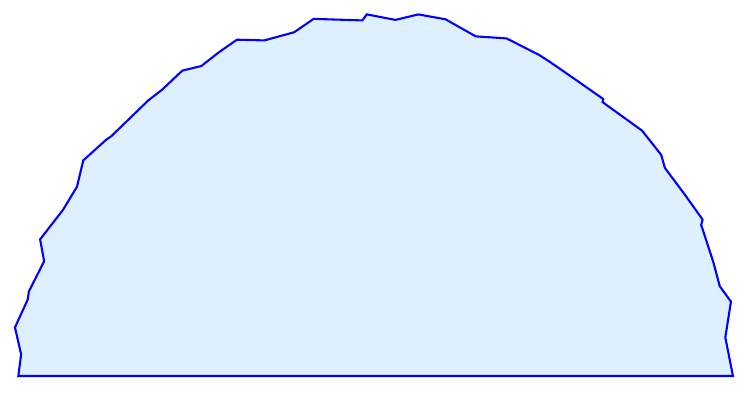}(a) \includegraphics[width=0.4\textwidth]{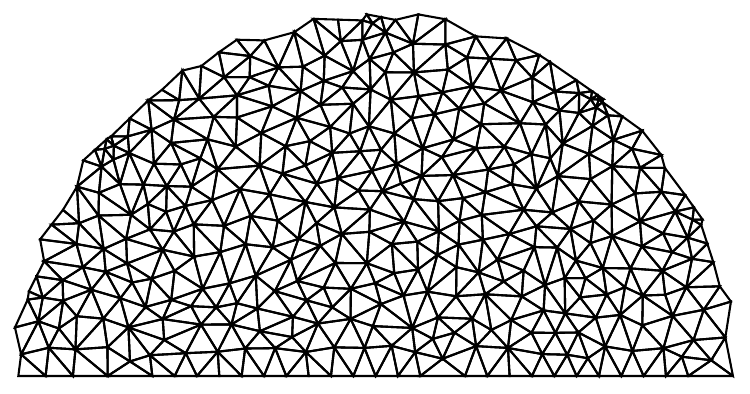}(b)
\caption{\label{fig3}A typical domain $\mathcal{S}$ (a), obtained with $N=45$ and $\gamma=0.02$, and one of its possible unstructured meshes (b) for the finite-element solver}
\end{figure}

The surface roughness of the microchannel curved boundary is modelled by assuming a random  change of the semicircle to a polygonal path, $\mathcal P$, with $N$ nodes having positions $(x_n , y_n)$ given by 
\eq{1}{
x_1 = r_0 \qc y_1 = 0 , \nn
x_n = \qty(r_0 + \delta r_n) \cos\theta_n \qc y_n = \qty(r_0 + \delta r_n) \sin\theta_n \qc \qfor n = 2,\ \ldots\ , N-1, \nn
x_N = -r_0 \qc y_N = 0, 
}
where $\delta r_n$ and $\theta_n$ are random variables within the ranges
\eq{2}{
-\gamma r_0 \le \delta r_n \le \gamma r_0 \qc \frac{(n - 2) \pi}{N-2} < \theta_n < \frac{(n-1) \pi}{N-2} .
}
The dimensionless parameter $\gamma$ employed in \eqref{2} is a measure of the roughness size relative to the nominal radius $r_0$. The polygonal path $\mathcal{P}$ is closed by joining the points $(x_N, y_N)$ and $(x_1,y_1)$ with the straight diametrical line. Thus, the closed polygonal path $\mathcal{P}$ is the boundary of the rough microchannel cross-section, denoted as $\mathcal{S}$. A sample region $\mathcal{S}$ obtained with $N=45$ and $\gamma=0.02$ is shown in Fig.~\ref{fig3}(a). 

The flow in the streamwise direction is stationary, incompressible and fully developed. Thus, the velocity field $\vb{u}$ has just one nonzero component, {\em i.e.}, that along the $z$ axis, denoted as $u$. As a consequence of the local mass balance equation, $\div\vb{u} = 0$, such a velocity component is independent of $z$. Hence, the average velocity across the microchannel is a constant denoted as $u_m$.

The governing equations for the fully developed flow (both hydrodynamically and thermally developed flow) are the local momentum balance and the local energy balance equations written as 
\eq{3}{
\laplacian{u} = \frac{1}{\mu}\, \dv{p}{z},\nn
u\, \pdv{T}{z} = \alpha \, \laplacian{T}  + \frac{\nu}{c}\, | \grad{u} |^2 ,
}
where $\dd p/\dd z$ is the constant pressure gradient driving the flow, $\mu$ is the dynamic viscosity, $\nu$ is the kinematic viscosity, $\alpha$ is the thermal diffusivity of the fluid and $c$ is its specific heat, while $T$ is the temperature field. The Laplacian $\laplacian$ operator and the gradient $\grad$ operator are intended as 2--dimensional in the $xy$ plane. Indeed, the fully developed regime implies that $\partial T/\partial z$ is a constant, hereafter denoted as $a$. As a consequence, we can write
\eq{4}{
\pdv{T}{z} = \dv{T_b}{z} = a , \nn
\qfor T_b = \frac{1}{\mathcal{S} u_m} \iint_{\mathcal{S}} T(x,y,z) u(x,y) \; \dd x\, \dd y  \qc u_m = \frac{1}{\mathcal{S}} \iint_{\mathcal{S}} u(x,y) \; \dd x\, \dd y,
}
with $T_b$ the bulk temperature. We emphasise that the term ${\nu}\, | \grad{u} |^2/{c}$ in the second equation \eqref{3} represents the viscous dissipation contribution to the local energy balance. 

For the sake of simplicity, we will always employ the same symbols $\mathcal{P}$ and $\mathcal{S}$ to denote both the geometrical objects and their measures (perimeter and area, respectively).
The solution of the first equation \eqref{3} with the no-slip boundary conditions 
\eq{5}{
u = 0 \qc (x,y) \in \mathcal{P},
}
yields the velocity profile $u(x,y)$ and, by employing \eqref{4}, the average velocity $u_m$. After having determined $u(x,y)$, the second equation \eqref{3} leads to the determination of $T(x,y,z)$. The solution process of the second equation \eqref{3} requires the statement of boundary conditions at the rough semicircular path, denoted as $\mathcal{P}_1$, and at the diametrical straight line, denoted as $\mathcal{P}_2$ (see Fig.~\ref{fig3}). Obviously, $\mathcal{P} = \mathcal{P}_1 \cup \mathcal{P}_2$. While the boundary condition at $\mathcal{P}_1$ prescribes thermal insulation,
\eq{6}{
\vu{n} \vdot \grad{T} = 0 \qc \qfor (x,y) \in \mathcal{P}_1 ,
}
with $\vu{n}$ the unit outward normal to $\mathcal{P}$, the thermal boundary condition at $\mathcal{P}_2$ can be formulated either as a {\sf H1} condition or as a {\sf H2} condition, where both are meant to describe a wall heating process,
\eq{7}{
\text{{\sf H1} condition:} \qquad \pdv{T}{x} = 0 \qc \qfor -r_0 < x < r_0 \qc y = 0;\nn
\text{{\sf H2} condition:} \qquad -k \pdv{T}{y} = q_w = \text{uniform} \qc \qfor -r_0 < x < r_0 \qc y = 0,
}
where $k$ is the fluid thermal conductivity.

The two different statements of the wall heating conditions, given by \eqref{7}, are those introduced and defined by \citet{shah1978laminar}. There are useful definitions of the average temperature of the heated boundary $\mathcal{P}_2$ and the average wall heat flux at $\mathcal{P}_2$,
\eq{8}{
T_w = \frac{1}{2 r_0} \int_{-r_0}^{r_0} T(x,0,z) \; \dd x \qc q_w = - \frac{k}{2 r_0} \int_{-r_0}^{r_0} \left. \pdv{T}{y} \right|_{y=0}  \dd x .
}
The difference between the boundary conditions {\sf H1} and {\sf H2} is that, in the former case, the temperature $T$ is uniform and equal to $T_w$ at every point in $\mathcal{P}_2$. On the other hand, with the condition {\sf H2}, the normal derivative $\partial T / \partial y$ is uniform over $\mathcal{P}_2$ and equal to $- q_w/k$. Both {\sf H1} and {\sf H2} wall heating conditions result in a nonzero value of the constant $a$, defined by \eqref{4}. Useful dimensionless parameters are the Poiseuille number and the Nusselt number defined as
\eq{9}{
\Po = \frac{u_0}{u_m}  \qc \Nu = \frac{2 r_h q_w}{k \qty(T_w - T_b)}  \qc
\qfor u_0 = - \frac{2 r_h^2}{\mu}\, \dv{p}{z},
}
where $r_h$ is the nominal hydraulic radius which, for a semicircular cross-section with radius $r_0$, is given by
\eq{10}{
r_h = \frac{\pi}{\pi + 2}\,r_0 .
}
By employing \eqref{4} and \eqref{8}, one can prove that both $\Po$ and $\Nu$ are constants, {\em i.e.}, they are independent of $z$.

\section{Nondimensional formulation}\label{nondim}
Let us define the dimensionless coordinates, operators and fields, denoted with asterisks,
\eq{11}{
\qty(x^*, y^*) =\frac{(x,y)}{r_h} \qc \grad^* = r_h \grad \qc {\nabla^*}^2 = r_h^2 \laplacian \qc u^* = \frac{u}{u_0} \qc T^* = k\, \frac{T - T_w}{q_w r_h}, 
}
as well as the Brinkman number and the dimensionless streamwise temperature gradient,
\eq{12}{
\Br = \frac{\mu\, u_m^2}{2 r_h q_w} \qc \sigma = \frac{k r_h u_m a}{\alpha\, q_w} .
}
The Brinkman number determines the strength of the viscous dissipation effect.
Using \eqref{4}, \eqref{11} and \eqref{12}, the governing equations \eqref{3} can be rewritten as
\eq{13}{
{\nabla^*}^2{u^*} + \frac{1}{2} = 0,\nn
{\nabla^*}^2 T^* - \Po\, \sigma u^* + 2\, \Po^2 \Br \, | \grad^*{\!u^*} |^2 = 0 .
}
The nondimensional boundary conditions \eqref{5}-\eqref{7} are given by
\eq{14}{
u^* = 0 \qc \qfor (x^*,y^*) \in \mathcal{P}^*,\nn
\vu{n} \vdot \grad^*{T^*} = 0 \qc \qfor (x^*,y^*) \in \mathcal{P}_1^* ,\nn
\text{{\sf H1} condition:} \qquad T^* = 0 \qc \qfor -\frac{\pi + 2}{\pi} < x^* < \frac{\pi + 2}{\pi} \qc y^* = 0;\nn
\text{{\sf H2} condition:} \qquad \pdv{T^*}{y^*} = -1 \qc \qfor -\frac{\pi + 2}{\pi} < x^* < \frac{\pi + 2}{\pi} \qc y^* = 0.
}
In fact, we denote with $\mathcal{S}^*$, $\mathcal{P}^*$ and $\mathcal{P}_1^*$ are the nondimensional objects obtained by scaling the coordinates $(x,y)$ as defined by \eqref{11}. In \eqref{14}, the {\sf H1} condition is reformulated by taking into account that $T$ is uniform over the diametrical side $\mathcal{P}_2$, so that $T=T_w$ at every point on that boundary.

On account of \eqref{4}, \eqref{9} and \eqref{11}, one can evaluate $\Po$ and $\Nu$ as
\eq{15}{
\Po = \frac{\mathcal{S}^*}{\displaystyle\iint_{\mathcal{S}^*} u^*(x^*, y^*)\; \dd x^*\, \dd y^*} \qc
\Nu = - \frac{2 \mathcal{S}^*}{\displaystyle \Po \iint_{\mathcal{S}^*} T^*(x^*, y^*)\, u^*(x^*, y^*) \; \dd x^*\, \dd y^*} .
}
As explicitly specified in \eqref{15}, $T^*$ is independent of $z^*$. This is a consequence of \eqref{11}, as well as of \eqref{4} and \eqref{8} which imply that $\partial T/\partial z = a = \dd T_w/\dd z$.

By integrating the second equation \eqref{13} over $\mathcal{S}^*$, by employing Gauss-Green theorem and by employing equations \eqref{4}, \eqref{8} and \eqref{11}, one obtains an expression for $\sigma$, namely
\eq{16}{
\sigma = \frac{2\qty(\pi + 2)}{\pi\, \mathcal{S}^*} + \frac{2 \Po^2 \Br}{\mathcal{S}^*} \iint_{\mathcal{S}^*} | \grad^*{\!u^*(x^*, y^*)} |^2\; \dd x^*\, \dd y^* = \frac{2\qty(\pi + 2)}{\pi\, \mathcal{S}^*} + \Po\, \Br , 
}
where the evaluation of the double integral has been carried by multiplying the first equation \eqref{13} by $u^*$, integrating by parts the resulting equation and then by employing the no-slip conditions together with \eqref{15}.
One may mention that a special case of the {\sf H1} condition is when $\sigma = 0$. This is a case such that the boundary temperature is not only uniform along the diametrical line $\mathcal{P}_2$, but also in the stramwise direction $z$. This special case defines the so-called {\sf T} condition, identified through a special value of $\Br$, given by
\eq{17}{
\Br_{\sf T} = - \frac{2\qty( \pi + 2 )}{\pi \, \Po\, \mathcal{S}^*} .
}
The solution of \eqref{13} and \eqref{14}, as well as the evaluation of $\Po$ and $\Nu$ based on \eqref{15}, can be obtained numerically by employing the finite-element PDE solver available in {\sl Wolfram 14} \citep{Mathematica}. The meshes employed for the solution process are unstructured, with triangular elements, and their generation is made through the built-in function {\tt ToElementMesh}. The refinement of the meshes employed is controlled via the parameter {\tt MaxCellMeasure} whose value defines the maximum area of the grid cells. Then, the numerical solution is developed by using the function {\tt NDSolveValue}, while the double integrals needed to compute $\Po$ and $\Nu$ are evaluated through function {\tt NIntegrate}. Figure~\ref{fig3}(b) shows a typical unstructured mesh for the sample domain $\mathcal{S}$ drawn in Fig.~\ref{fig3}(a) for $N=45$ and $\gamma = 0.02$. 

\begin{table}
\centering
\begin{tabular}{l l l}
\hline\\[-12pt]
{\tt MaxCellMeasure} & $\Po$ & $|\text{relative error}|$ (\%)\\
\hline\\[-10pt]
$10^{-1}$ & 15.7686 & 0.011\\
$10^{-1.5}$ & 15.7672 & 0.0024\\
$10^{-2}$ & 15.7669 & 0.00039\\
$10^{-2.5}$ & 15.7668 & 0.000030\\
$10^{-3}$ & 15.7668 & $4.0 \times 10^{-6}$\\
$10^{-3.5}$ & 15.7668 & $5.3 \times 10^{-7}$\\
$10^{-4}$ & 15.7668 & $1.4 \times 10^{-8}$\\
\hline
\end{tabular}
\caption{\label{tab1}Test of the numerical finite-element solver: evaluation of $\Po$ with $\gamma \to 0$ and different mesh refinements ({\tt MaxCellMeasure}) and the absolute value of the relative error with respect to the exact result \eqref{18}}
\end{table}

\section{The perfectly smooth semicircular duct}\label{smoothscd}
An important benchmark case is when the surface roughness tends to zero $(\gamma \to 0)$ \citep{shah1978laminar, manglik1988laminar, LANGURI2011139, alassar2014fully, mukherjee2017laminar, kyritsi2018newtonian}. In this case, there exists an analytical expression of $\Po$ available in the literature \citep{LANGURI2011139, kyritsi2018newtonian},
\eq{18}{
\Po = \frac{8 \pi^4}{\qty(\pi + 2)^2 \qty(\pi^2 - 8)} \approx 15.7668 .
}
This exact result is used to test the convergence of the numerical solver by defining meshes with a decreasing value of the parameter {\tt MaxCellMeasure}. The results of this test are reported in Table~\ref{tab1}. This table shows that an excellent agreement within the first six significant figures is already attained with ${\tt MaxCellMeasure \to 10^{-2.5}}$. Hereafter, all the computations for the microchannels with a rough or smooth boundary will be carried out with ${\tt MaxCellMeasure \to 10^{-3}}$.

In the case where $\gamma \to 0$, we have
\eq{19}{
\mathcal{S}^* = \frac{\qty(\pi + 2)^2}{2 \pi} \qc
\sigma = \frac{4}{\pi + 2 } + \frac{8 \pi^4\, \Br}{(\pi + 2)^2 \left(\pi^2-8\right)} \qc
\Br_{\sf T} = - \frac{\qty( \pi + 2 ) \qty(\pi^2 - 8)}{2 \pi^4} ,
}
where the expressions of $\sigma$ and $\Br_{\sf T}$ have been obtained by using \eqref{16}-\eqref{18}. On the other hand, the values of $\Nu$ for the cases {\sf H1} and {\sf H2} are obtained for prescribed values of $\Br$. Except for the {\sf T} condition where $\Br = \Br_{\sf T}$, we will consider only non-negative values of $\Br$ as the scope of our analysis is the case of a heated diametrical boundary.

\begin{table}
\centering
\begin{tabular}{l l l}
\hline\\[-12pt]
$\Br$ & $\Nu_{\sf H1}$ & $\Nu_{\sf H2}$ \\
\hline\\[-10pt]
0 & 3.28168 & 3.17683 \\
0.01 & 3.17278 & 3.07468 \\
0.1 & 2.44317 & 2.38458 \\
0.5 & 1.20826 & 1.19376 \\
1 & 0.740440 & 0.734967 \\
1.5 & 0.533771 & 0.530922 \\
2 & 0.417297 & 0.415553 \\
2.5 & 0.342549 & 0.341373 \\
3 & 0.290512 & 0.289666 \\
3.5 & 0.252200 & 0.251562\\
4 & 0.222815 & 0.222317\\
4.5 & 0.199563 & 0.199164 \\
5 & 0.180706 & 0.180378 \\
\hline
\end{tabular}
\caption{\label{tab2}Values of $\Nu_{\sf H1}$ and $\Nu_{\sf H2}$ with $\gamma \to 0$ for different Brinkman numbers}
\end{table}

\begin{figure}
\centering
\includegraphics[height=0.4\textwidth]{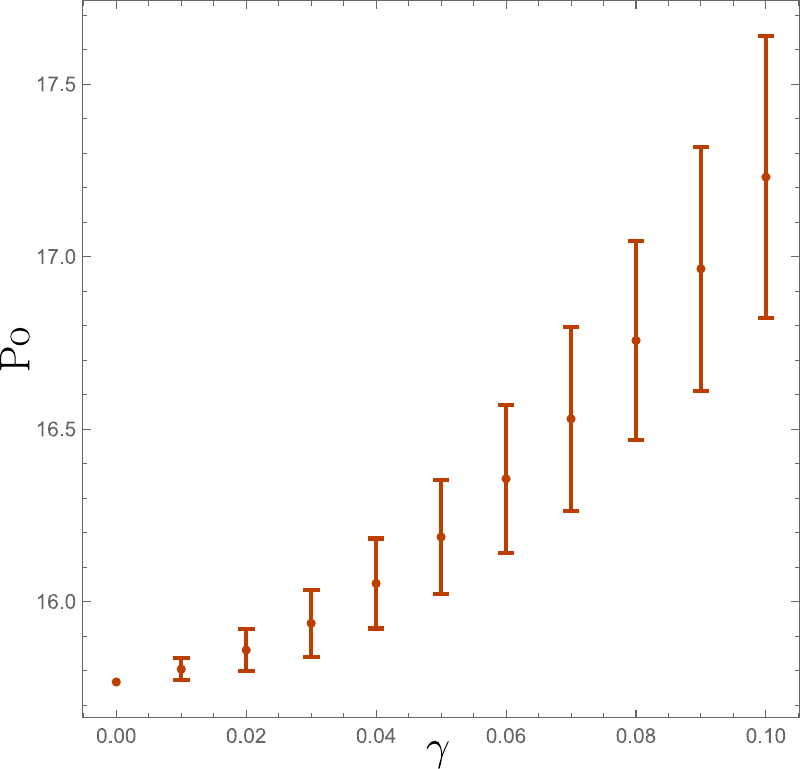}(a) \includegraphics[height=0.4\textwidth]{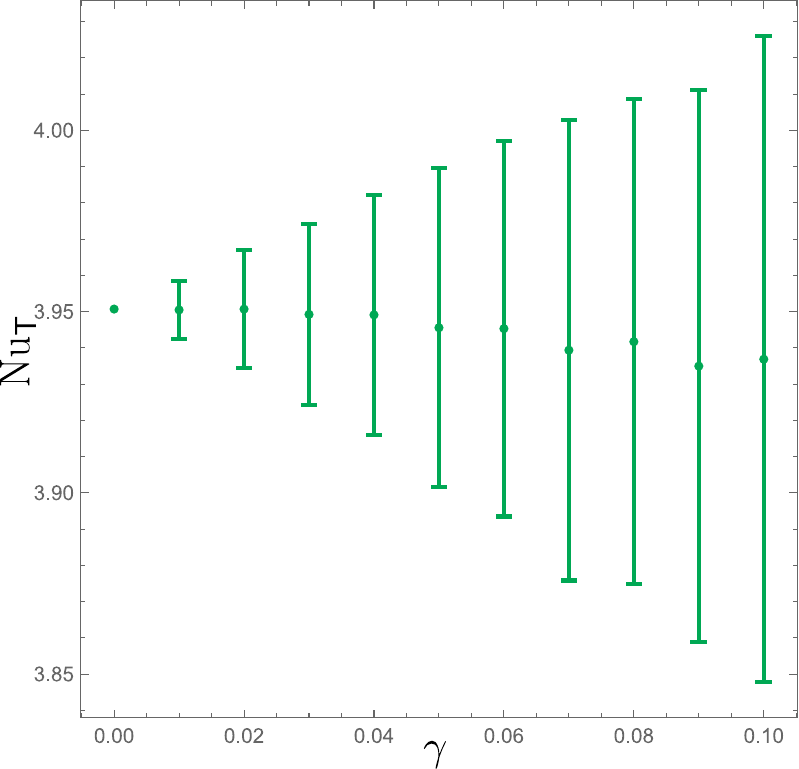}(b)
\caption{\label{fig4}Plots of $\Po$ (a) and $\Nu_{\sf T}$ (b) versus $\gamma$. The dots indicate the mean value over the statistical population, while the segments show the uncertainties in a one standard deviation range from the mean value}
\end{figure}

\begin{figure}[ht!]
\centering
\includegraphics[height=0.4\textwidth]{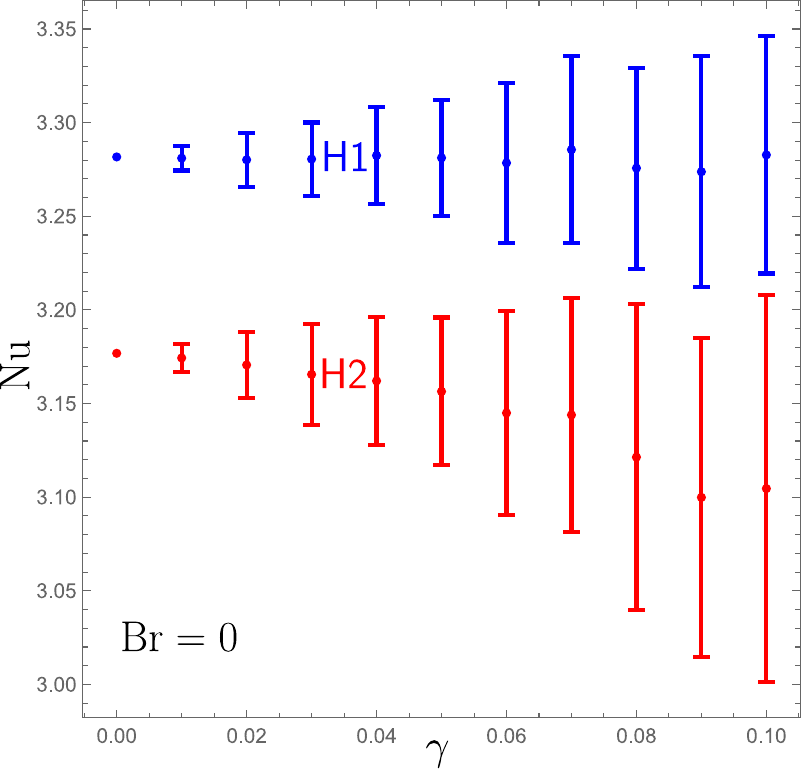}(a) \includegraphics[height=0.4\textwidth]{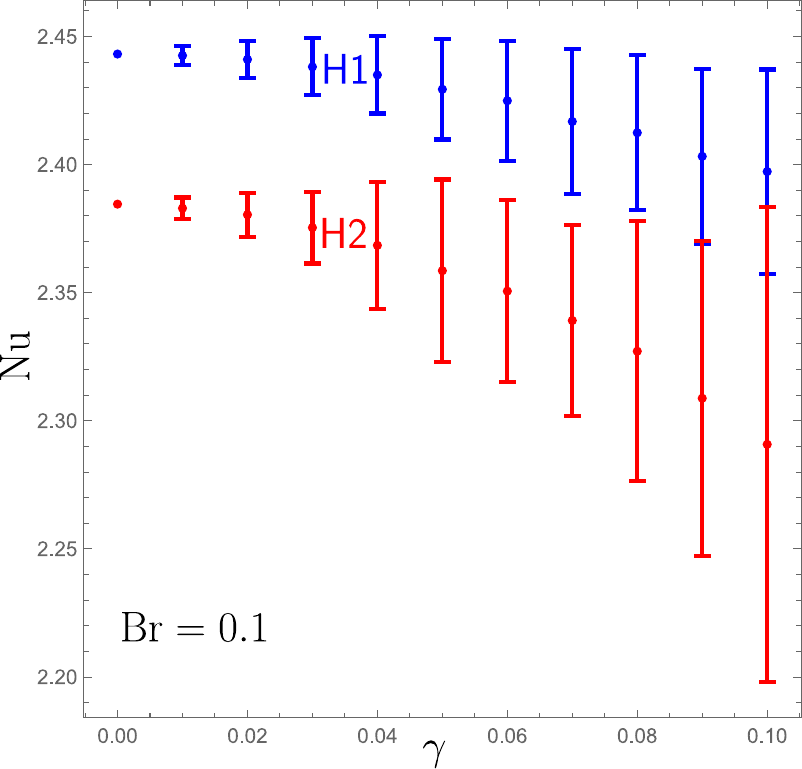}(b)\\
\includegraphics[height=0.4\textwidth]{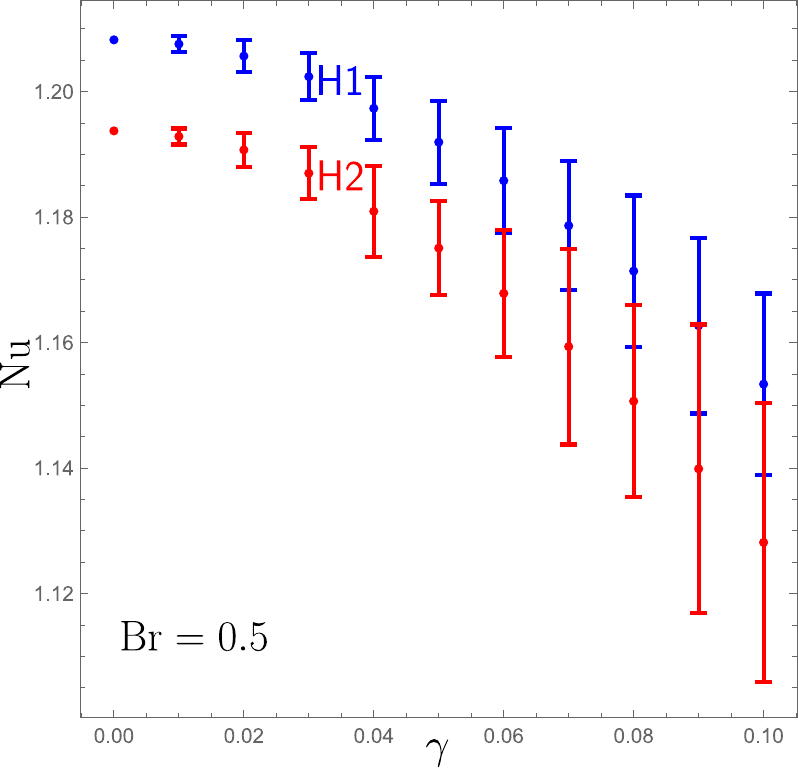}(c) \includegraphics[height=0.4\textwidth]{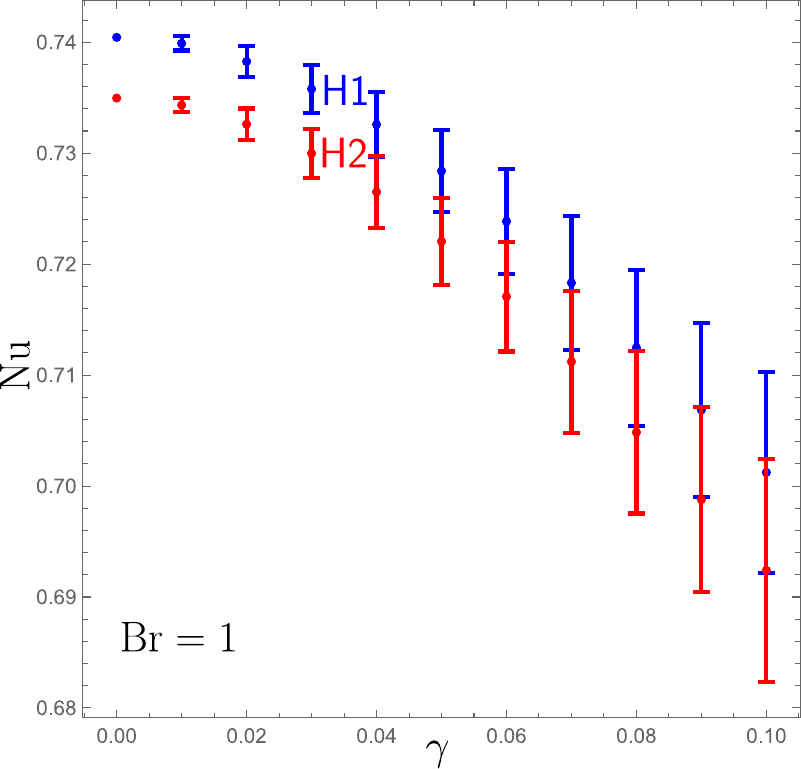}(d)
\caption{\label{fig5}Plots of $\Nu_{\sf H1}$ (in blue) and $\Nu_{\sf H2}$ (in red) versus $\gamma$ for $\Br=0$ (a), $\Br = 0.1$(b), $\Br=0.5$ (c) and $\Br=1$ (d). The dots indicate the mean value over the statistical population, while the segments show the uncertainties in a one standard deviation range from the mean value}
\end{figure}

The values of $\Nu$ can be determined numerically for the {\sf T} boundary condition, as well as for the {\sf H1} and {\sf H2} boundary conditions with fixed values of $\Br$. The value of $\Nu_{\sf T}$ is
\eq{20}{
\Nu_{\sf T} = 3.95071 ,
}
while values of $\Nu_{\sf H1}$ and $\Nu_{\sf H2}$ are reported in Table~\ref{tab2} for some Brinkman numbers. An evident feature of these data is that $\Nu_{\sf H1}$ is greater than $\Nu_{\sf H2}$ as typical of fully developed forced convection in ducts \citep{shah1978laminar}. However, the relative discrepancy is always small and it decreases with $\Br>0$ being around $3.3\%$, at its largest, when $\Br = 0$.

\begin{figure}[p]
\centering
\includegraphics[height=0.4\textwidth]{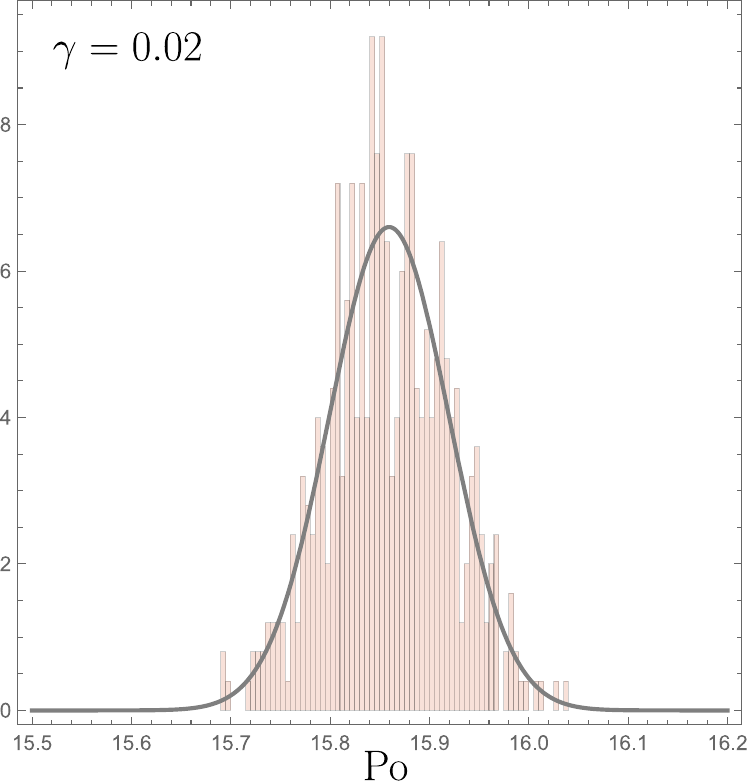}(a) \includegraphics[height=0.4\textwidth]{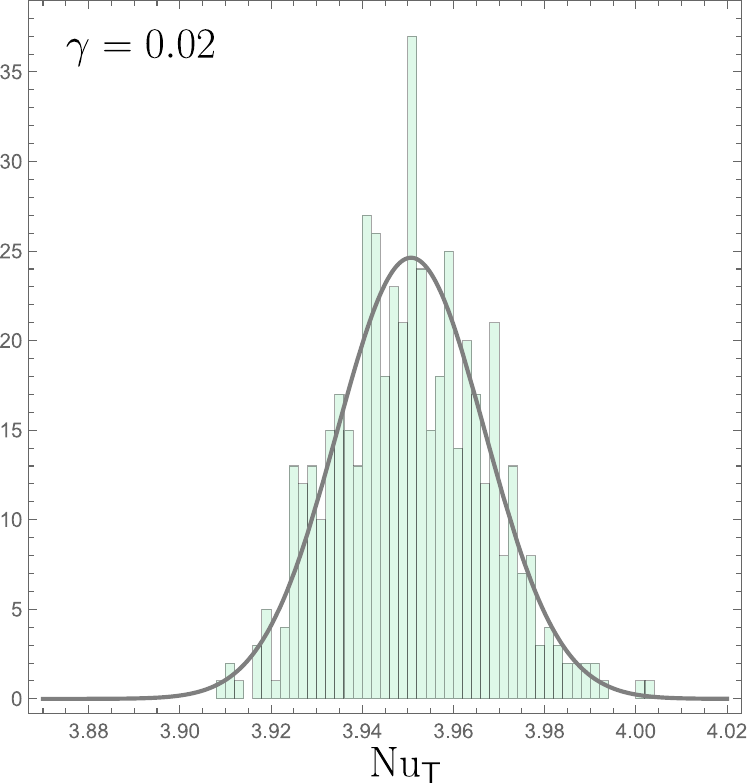}(b)\\
\includegraphics[height=0.4\textwidth]{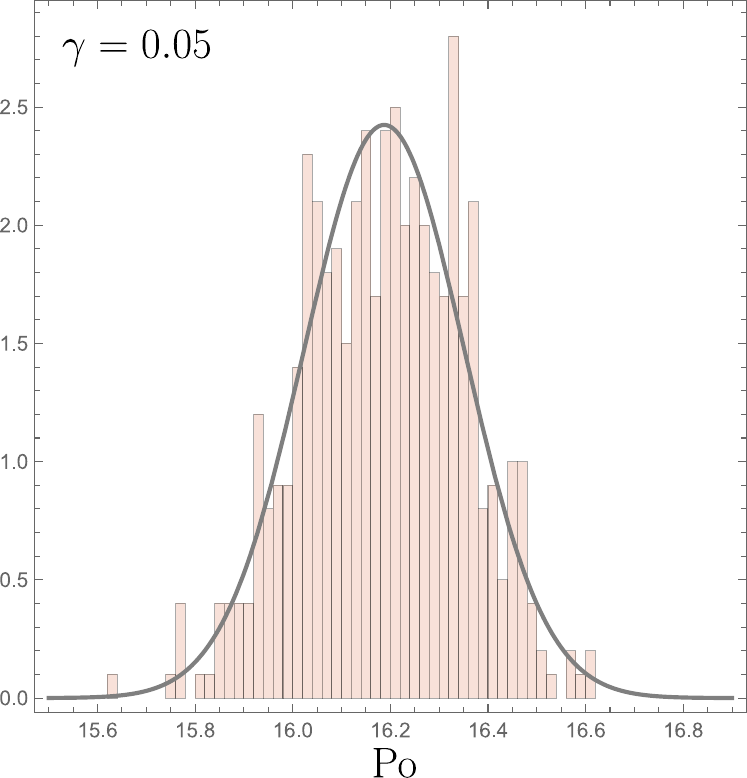}(c) \includegraphics[height=0.4\textwidth]{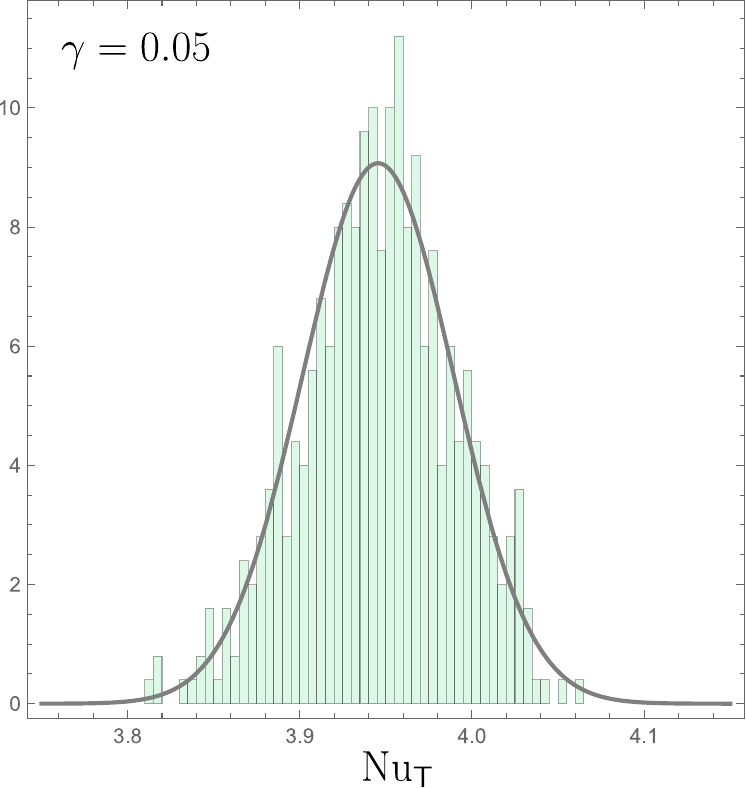}(d)\\
\includegraphics[height=0.4\textwidth]{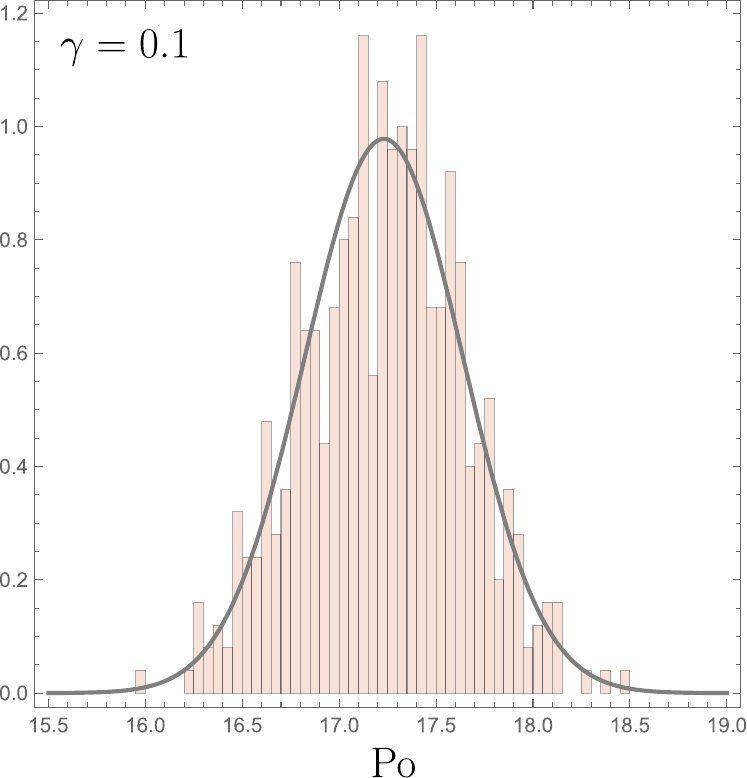}(e) \includegraphics[height=0.4\textwidth]{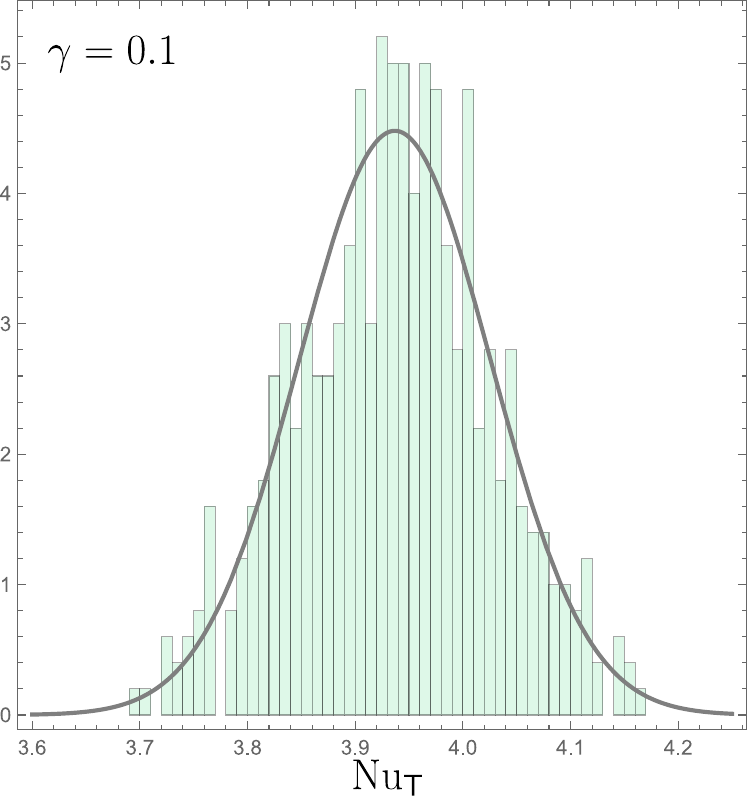}(f)
\caption{\label{fig6}Probability density plots with the actual data for $\Po$ and $\Nu_{\sf T}$ for $\gamma = 0.02$ (a, b), $\gamma=0.05$ (c, d) and $\gamma=0.1$ (e,f). In each frame, the gray line shows the normal distribution with the same mean value and standard deviation of the data}
\end{figure}

\section{Discussion of the effects of wall roughness}\label{diswalrou}
The gradual increase of the wall roughness is monitored by fixing values of $\gamma$ within the range $0 < \gamma \le 0.1$. For each value of $\gamma$, a population of $500$ different cross-sections is generated by means of random polygonal paths defined through equations \eqref{1} and \eqref{2}. In every population, we evaluate the mean value and the standard deviation for the parameters $\Po$, $\Nu_{\sf T}$, $\Nu_{\sf H1}$ and $\Nu_{\sf H2}$. The first two parameters $\Po$ and $\Nu_{\sf T}$ are not affected by $\Br$. This feature already emerged in the case of the smooth semicircular duct discussed in Section~\ref{smoothscd}. In fact, it is a general feature valid for every $\gamma$ as $\Po$ is determined only by solving the local momentum balance equation, which is independent of $\Br$, with no-slip boundary conditions.
Furthermore, $\Nu_{\sf T}$ is obtained by assigning a precise Brinkman number, namely $\Br = \Br_{\sf T}$. On the other hand, $\Nu_{\sf H1}$ and $\Nu_{\sf H2}$ depend on  the value of the Brinkman number, exactly as it happens in the case of the smooth semicircular duct (see Table~\ref{tab2}).

Figure~\ref{fig4} shows the effect of roughness on the Poiseuille number, $\Po$, and on the Nusselt number for the $\sf T$ boundary condition, $\Nu_{\sf T}$. In fact, Fig.~\ref{fig4}(b) reveals that there is no statistically relevant influence of roughness on $\Nu_{\sf T}$, at least as far as $\gamma \le 0.1$, if not for a tendency to increasing uncertainty of the data as $\gamma$ increases. Anyway, the relative uncertainty is approximately $2.3\%$ at its largest (for $\gamma = 0.1$). On the other hand, Fig.~\ref{fig4}(a) shows that $\Po$ increases with $\gamma$.  For this parameter, the relative uncertainty is, in every case, lower than approximately $2.4\%$. The roughness-induced increase of $\Po$ is caused by the gradually increasing hydraulic resistance when the average length of surface spikes over the semicircular boundary becomes larger and larger. In dimensional terms, this effect corresponds to a gradually increasing value of $\dd p/\dd z$ for a given $u_m$, as one can easily infer from \eqref{9}.

Figure~\ref{fig5} displays the mean values and standard deviations for the Nusselt number versus $\gamma$ relative to the boundary conditions {\sf H1} and {\sf H2}. In this figure, frames (a) to (d) address increasing values of the Brinkman number from $0$ to $1$. An evident effect of the raising importance of the viscous dissipation effect is that it triggers an effect of the growing value of $\gamma$. In particular, with $\Br = 0$ (negligible viscous dissipation) and $\Br = 0.1$, namely for the plots shown in Fig.~\ref{fig5}(a) and  Fig.~\ref{fig5}(b), the increasing roughness does not have a statistically significant effect on $\Nu$ for both {\sf H1} and {\sf H2} conditions. On the other hand, such an effect emerges for $\Br = 0.5$ and $\Br = 1$, in Fig.~\ref{fig5}(c) and  Fig.~\ref{fig5}(d), as an increasing $\gamma$ determines a weak, though indisputable, decrease in the Nusselt number in the two cases {\sf H1} and {\sf H2}. The Nusselt number for the {\sf H1} condition turns out to be greater than that for the  {\sf H2} condition, even though the relatively large dispersion of the data may determine a statistically relevant overlapping especially at larger values of $\gamma$ and $\Br$. This special interplay between wall roughness and viscous dissipation effect is physically grounded. In fact, for a fixed value of $\Br$, viscous heating is enhanced by a larger roughness as an increase in $\gamma$ means larger local values of the source term $2\, \Po^2 \Br \, | \grad^*{\!u^*} |^2$ in the energy balance \eqref{13}, as induced both by the larger value of $\Po$ and by the typically larger local values of the gradient of $u^*$. The latter feature is an expected consequence of the spikes and irregularities of the roughly semicircular part of the microchannel boundary. 

If we compare the results reported in Figs.~\ref{fig4} and \ref{fig5} with those presented in \citet{barletta2025shape}, we recognise that the main qualitative discrepancy emerges from the effects of the surface roughness on the parameter $\Nu_{\sf T}$. In fact, \citet{barletta2025shape} found that, in a circular rough microchannel, $\Nu_{\sf T}$ displays a marked decrease with the roughness parameter. The main reason for the discrepancy with the trend shown in Fig.~\ref{fig4}(b) is that the heat exchange surface (the diametrical side of the channel) is smooth in the present study, while it is affected by roughness in \citet{barletta2025shape}. We emphasise that, in the latter paper, a microduct with a nominally circular cross-section is studied where the whole side boundary is considered as rough and subject to heat transfer. Another correlated aspect of the comparison is that the effects of roughness on the parameters $\Nu_{\sf H1}$ and $\Nu_{\sf H2}$ are evident in Fig.~\ref{fig5}, but they are definitely smaller than those found by \citet{barletta2025shape}. Furthermore, $\Nu_{\sf H1}$ and $\Nu_{\sf H2}$ are affected by the surface roughness, according to Fig.~\ref{fig5}, only if viscous dissipation is taken into account and if it is sufficiently intense. On the other hand, roughness effects on $\Nu_{\sf H1}$ and $\Nu_{\sf H2}$ are found in \citet{barletta2025shape} also when $\Br \to 0$.  Thus, we may conclude that the effects of surface roughness, at the microchannel boundary, on the heat transfer rate are more relevant when the surface from which heat is actually transferred is rough.

A closer view into the statistical distribution of data for the Poiseuille number and the Nusselt number with the {\sf T} boundary condition is provided in Fig.~\ref{fig6}. In this figure, the probability density distributions for the 500 microchannels populations are displayed via histograms compared with the normal distributions having the same mean value and standard deviation of the actual data. The different frames correspond to different values of the roughness parameter $\gamma$, namely: $0.02$ (frames a and b), $0.05$ (frames c and d) and $0.1$ (frames e and f).
Figure~\ref{fig6} suggests that the normal distribution represents a reasonable model for the computed data even if larger populations would be needed to attain a visually satisfactory comparison. We stress that 500 elements in each population is a statistically reliable target which, besides, shows up acceptable computational times for the finite-element solution procedure.

\section{Conclusions}
A numerical analysis of the laminar forced convection in semicircular microchannels has been carried out in order to gain information on the thermal characteristics of heat sinks designed for cooling a heated metal plate. While the metal plate, corresponding to the diametrical boundary of the microchannel has been supposed to be polished and, hence, perfectly smooth, the semicircular boundary has been considered as having a significant roughness. 

The plane diametrical boundary is heated and the heating has been modelled through either a {\sf H1} or a {\sf H2} boundary condition, consistently with the definitions given by \citet{shah1978laminar}. A special case of the {\sf H1} boundary condition, {\em i.e.} the {\sf T} boundary condition, has been also studied for the sake of completeness as it represents the case where the diametrical boundary is isothermal both in the perimetral and in the streamwise directions. The analysis of the special {\sf T} boundary condition shows up an independence from the Brinkman number or, in other terms, the Brinkman number is fixed to a geometry-dependent value, $\Br_{\sf T}$.

Compared to the heat flux supplied through the diametrical boundary, the semicircular boundary heat flux is considerably smaller so that, in this analysis, such a boundary has been considered as thermally insulated.

A further internal source of fluid heating is viscous dissipation, namely the mechanical work to heat conversion due to the forced flow in the microchannel. Such an effect, which may be significant especially in the small scale typical of microfluidics \citep{bruus2007theoretical} is parametrised by the Brinkman number, $\Br$. Surface roughness along the semicircular boundary has been described by the dimensionless parameter $\gamma$, defined as the maximum ratio between the length of the surface spikes to the semicircle radius. The computational approach to the surface roughness effects has been grounded on a statistical scheme by generating, in each case, a population of 500 randomly generated deformations of the semicircle. Each deformed path is a polygonal connecting 45 points, defined within the tolerance introduced by the roughness parameter $\gamma$. The numerical solution of the local momentum balance and the local energy balance equations has been obtained for each randomly generated domain by employing the finite element method developed within the software environment {\sl Wolfram 14} \citep{Mathematica}.
The main results obtained in our study can be summarized as follows:
\begin{itemize}
\item The flow mechanical resistance, described by the Poiseuille number $\Po$, turns out to increase with the wall roughness parameter $\gamma$ displaying an increasing statistical dispersion of the data and, hence, an increasing standard deviation, as the roughness increases.
\item The Nusselt number for the special {\sf T} boundary condition turned out to be unaffected by the wall roughness even though its standard deviation increases with $\gamma$.
\item The Nusselt number $\Nu$ for the cases {\sf H1} and {\sf H2} is independent of $\gamma$ when the viscous dissipation effect is neglected $(\Br \to 0)$, though $\Nu_{\sf H1}$ is generally larger than $\Nu_{\sf H2}$. The latter is a typical feature of laminar forced convection flows \citep{shah1978laminar}.
\item When viscous dissipation is non-negligible, then both $\Nu_{\sf H1}$ and $\Nu_{\sf H2}$ tend to decrease with $\gamma$, showing up a significant increase of the standard deviation which, in some cases, tends to yield a partial overlap of the confidence intervals for these parameters.
\end{itemize}
By comparing the results of the present analysis with those reported in a similar study relative to a microduct with a nominally circular cross-section \citep{barletta2025shape}, we have concluded that the effects of roughness are definitely minor when the heat transfer is from a perfectly smooth part of the boundary, even if the rest of the boundary is affected by roughness.

In our analysis, the special case of the smooth semicircular duct $(\gamma \to 0)$ has been considered as a benchmark case which is well-know from the existing literature only with reference to the Poiseuille number \citep{LANGURI2011139, kyritsi2018newtonian}. On the other hand, the values of the Nusselt number have been obtained numerically in this paper, due to the peculiar assignment of the thermal boundary conditions which differs from those typically considered in the literature.

We emphasise that the temperature boundary condition of thermal insulation at the nominally semicircular boundary can be improved to a more realistic one. A possible candidate is a third-kind boundary condition allowing for a finite thermal resistance of the slab where the semicircular passages are etched. A significant improvement of this study can be achieved by extending the analysis to a three-dimensional model where the roughness is not streamwise invariant. We mention that, although a three-dimensional study is significantly more realistic, its implementation is by and large more demanding in terms of computational time and in terms of number of governing parameters. The Reynolds number and the Prandtl number would enter the governing equations making any attempt to generality in the discussion of the results a considerably difficult, though challenging, task.

\section*{Acknowledgements}
The work by Antonio Barletta, Michele Celli and Pedro Vayssi\`ere Brand\~ao was supported by Alma Mater Studiorum Universit\`a di Bologna, grant number
RFO-2024.


\begin{thebibliography}{34}
\expandafter\ifx\csname natexlab\endcsname\relax\def\natexlab#1{#1}\fi
\providecommand{\bibinfo}[2]{#2}
\ifx\xfnm\relax \def\xfnm[#1]{\unskip,\space#1}\fi
\bibitem[{Khan and Fartaj(2011)}]{khan2011review}
\bibinfo{author}{M.~G. Khan}, \bibinfo{author}{A.~Fartaj},
\newblock \bibinfo{title}{A review on microchannel heat exchangers and
  potential applications},
\newblock \bibinfo{journal}{International Journal of Energy Research}
  \bibinfo{volume}{35} (\bibinfo{year}{2011}) \bibinfo{pages}{553--582}.
\bibitem[{Kandlikar(2012)}]{kandlikar2012history}
\bibinfo{author}{S.~G. Kandlikar},
\newblock \bibinfo{title}{History, advances, and challenges in liquid flow and
  flow boiling heat transfer in microchannels: {A} critical review},
\newblock \bibinfo{journal}{ASME Journal of Heat Transfer}
  \bibinfo{volume}{134} (\bibinfo{year}{2012}) \bibinfo{pages}{034001}.
\bibitem[{Siddiqui and Zubair(2017)}]{siddiqui2017efficient}
\bibinfo{author}{O.~K. Siddiqui}, \bibinfo{author}{S.~M. Zubair},
\newblock \bibinfo{title}{Efficient energy utilization through proper design of
  microchannel heat exchanger manifolds: {A} comprehensive review},
\newblock \bibinfo{journal}{Renewable and Sustainable Energy Reviews}
  \bibinfo{volume}{74} (\bibinfo{year}{2017}) \bibinfo{pages}{969--1002}.
\bibitem[{Gilmore et~al.(2018)Gilmore, Timchenko, and
  Menictas}]{gilmore2018microchannel}
\bibinfo{author}{N.~Gilmore}, \bibinfo{author}{V.~Timchenko},
  \bibinfo{author}{C.~Menictas},
\newblock \bibinfo{title}{Microchannel cooling of concentrator photovoltaics:
  {A} review},
\newblock \bibinfo{journal}{Renewable and Sustainable Energy Reviews}
  \bibinfo{volume}{90} (\bibinfo{year}{2018}) \bibinfo{pages}{1041--1059}.
\bibitem[{Deng et~al.(2021)Deng, Zeng, and Sun}]{deng2021review}
\bibinfo{author}{D.~Deng}, \bibinfo{author}{L.~Zeng}, \bibinfo{author}{W.~Sun},
\newblock \bibinfo{title}{A review on flow boiling enhancement and fabrication
  of enhanced microchannels of microchannel heat sinks},
\newblock \bibinfo{journal}{International Journal of Heat and Mass Transfer}
  \bibinfo{volume}{175} (\bibinfo{year}{2021}) \bibinfo{pages}{121332}.
\bibitem[{Harris et~al.(2022)Harris, Wu, Zhang, and
  Angelopoulou}]{harris2022overview}
\bibinfo{author}{M.~Harris}, \bibinfo{author}{H.~Wu},
  \bibinfo{author}{W.~Zhang}, \bibinfo{author}{A.~Angelopoulou},
\newblock \bibinfo{title}{Overview of recent trends in microchannels for heat
  transfer and thermal management applications},
\newblock \bibinfo{journal}{Chemical Engineering and Processing -- Process
  Intensification} \bibinfo{volume}{181} (\bibinfo{year}{2022})
  \bibinfo{pages}{109155}.
\bibitem[{Heidarshenas et~al.(2024)Heidarshenas, El-Shafay, Sajadi, and
  Yuan}]{heidarshenas2024enhancing}
\bibinfo{author}{B.~Heidarshenas}, \bibinfo{author}{A.~El-Shafay},
  \bibinfo{author}{S.~M. Sajadi}, \bibinfo{author}{Y.~Yuan},
\newblock \bibinfo{title}{Enhancing efficiency in microscale systems with
  microchannels: {A} review},
\newblock \bibinfo{journal}{Journal of Thermal Analysis and Calorimetry}
  (\bibinfo{year}{2024}) \bibinfo{pages}{1--26}.
\bibitem[{Dash et~al.(2022)Dash, Nanda, and Rout}]{dash2022role}
\bibinfo{author}{B.~Dash}, \bibinfo{author}{J.~Nanda}, \bibinfo{author}{S.~K.
  Rout},
\newblock \bibinfo{title}{The role of microchannel geometry selection on heat
  transfer enhancement in heat sinks: {A} review},
\newblock \bibinfo{journal}{Heat Transfer} \bibinfo{volume}{51}
  (\bibinfo{year}{2022}) \bibinfo{pages}{1406--1424}.
\bibitem[{Yu et~al.(2024)Yu, Li, and Cao}]{yu2024comprehensive}
\bibinfo{author}{Z.-Q. Yu}, \bibinfo{author}{M.-T. Li}, \bibinfo{author}{B.-Y.
  Cao},
\newblock \bibinfo{title}{A comprehensive review on microchannel heat sinks for
  electronics cooling},
\newblock \bibinfo{journal}{International Journal of Extreme Manufacturing}
  \bibinfo{volume}{6} (\bibinfo{year}{2024}) \bibinfo{pages}{022005}.
\bibitem[{Dong et~al.(2024)Dong, Zhang, Liu, Wang, and Fang}]{DONG2024125001}
\bibinfo{author}{W.~Dong}, \bibinfo{author}{X.~Zhang},
  \bibinfo{author}{B.~Liu}, \bibinfo{author}{B.~Wang},
  \bibinfo{author}{Y.~Fang},
\newblock \bibinfo{title}{Research progress on passive enhanced heat transfer
  technology in microchannel heat sink},
\newblock \bibinfo{journal}{International Journal of Heat and Mass Transfer}
  \bibinfo{volume}{220} (\bibinfo{year}{2024}) \bibinfo{pages}{125001}.
\bibitem[{Xu et~al.(2002)Xu, Ooi, Mavriplis, and Zaghloul}]{BXu2003}
\bibinfo{author}{B.~Xu}, \bibinfo{author}{K.~T. Ooi},
  \bibinfo{author}{C.~Mavriplis}, \bibinfo{author}{M.~E. Zaghloul},
\newblock \bibinfo{title}{Evaluation of viscous dissipation in liquid flow in
  microchannels},
\newblock \bibinfo{journal}{Journal of Micromechanics and Microengineering}
  \bibinfo{volume}{13} (\bibinfo{year}{2002}) \bibinfo{pages}{53}.
\bibitem[{Koo and Kleinstreuer(2004)}]{KOO20043159}
\bibinfo{author}{J.~Koo}, \bibinfo{author}{C.~Kleinstreuer},
\newblock \bibinfo{title}{Viscous dissipation effects in microtubes and
  microchannels},
\newblock \bibinfo{journal}{International Journal of Heat and Mass Transfer}
  \bibinfo{volume}{47} (\bibinfo{year}{2004}) \bibinfo{pages}{3159--3169}.
\bibitem[{Croce and D’Agaro(2004)}]{CROCE2004601}
\bibinfo{author}{G.~Croce}, \bibinfo{author}{P.~D’Agaro},
\newblock \bibinfo{title}{Numerical analysis of roughness effect on microtube
  heat transfer},
\newblock \bibinfo{journal}{Superlattices and Microstructures}
  \bibinfo{volume}{35} (\bibinfo{year}{2004}) \bibinfo{pages}{601--616}.
\bibitem[{Croce and D'Agaro(2005)}]{croce2005numerical}
\bibinfo{author}{G.~Croce}, \bibinfo{author}{P.~D'Agaro},
\newblock \bibinfo{title}{Numerical simulation of roughness effect on
  microchannel heat transfer and pressure drop in laminar flow},
\newblock \bibinfo{journal}{Journal of Physics D: Applied Physics}
  \bibinfo{volume}{38} (\bibinfo{year}{2005}) \bibinfo{pages}{1518--1530}.
\bibitem[{Bruus(2007)}]{bruus2007theoretical}
\bibinfo{author}{H.~Bruus}, \bibinfo{title}{Theoretical Microfluidics},
  \bibinfo{publisher}{Oxford University Press}, \bibinfo{year}{2007}.
\bibitem[{Croce et~al.(2007)Croce, D'Agaro, and Nonino}]{croce2007three}
\bibinfo{author}{G.~Croce}, \bibinfo{author}{P.~D'Agaro},
  \bibinfo{author}{C.~Nonino},
\newblock \bibinfo{title}{Three-dimensional roughness effect on microchannel
  heat transfer and pressure drop},
\newblock \bibinfo{journal}{International Journal of Heat and Mass Transfer}
  \bibinfo{volume}{50} (\bibinfo{year}{2007}) \bibinfo{pages}{5249--5259}.
\bibitem[{Van~Rij et~al.(2009)Van~Rij, Ameel, and Harman}]{van2009effect}
\bibinfo{author}{J.~Van~Rij}, \bibinfo{author}{T.~Ameel},
  \bibinfo{author}{T.~Harman},
\newblock \bibinfo{title}{The effect of viscous dissipation and rarefaction on
  rectangular microchannel convective heat transfer},
\newblock \bibinfo{journal}{International Journal of Thermal Sciences}
  \bibinfo{volume}{48} (\bibinfo{year}{2009}) \bibinfo{pages}{271--281}.
\bibitem[{Turgay and Yazicioglu(2009)}]{turgay2009effect}
\bibinfo{author}{M.~B. Turgay}, \bibinfo{author}{A.~G. Yazicioglu},
\newblock \bibinfo{title}{Effect of surface roughness in parallel-plate
  microchannels on heat transfer},
\newblock \bibinfo{journal}{Numerical Heat Transfer, Part A: Applications}
  \bibinfo{volume}{56} (\bibinfo{year}{2009}) \bibinfo{pages}{497--514}.
\bibitem[{Nonino et~al.(2010)Nonino, {\relax Del Giudice}, and
  Savino}]{Nonino2010}
\bibinfo{author}{C.~Nonino}, \bibinfo{author}{S.~{\relax Del Giudice}},
  \bibinfo{author}{S.~Savino},
\newblock \bibinfo{title}{Temperature-dependent viscosity and viscous
  dissipation effects in microchannel flows with uniform wall heat flux},
\newblock \bibinfo{journal}{Heat Transfer Engineering} \bibinfo{volume}{31}
  (\bibinfo{year}{2010}) \bibinfo{pages}{682--691}.
\bibitem[{Pelevi{\'c} and van~der Meer(2016)}]{pelevic2016heat}
\bibinfo{author}{N.~Pelevi{\'c}}, \bibinfo{author}{T.~H. van~der Meer},
\newblock \bibinfo{title}{Heat transfer and pressure drop in microchannels with
  random roughness},
\newblock \bibinfo{journal}{International Journal of Thermal Sciences}
  \bibinfo{volume}{99} (\bibinfo{year}{2016}) \bibinfo{pages}{125--135}.
\bibitem[{Mukherjee et~al.(2017)Mukherjee, Biswal, Chakraborty, and
  DasGupta}]{mukherjee2017effects}
\bibinfo{author}{S.~Mukherjee}, \bibinfo{author}{P.~Biswal},
  \bibinfo{author}{S.~Chakraborty}, \bibinfo{author}{S.~DasGupta},
\newblock \bibinfo{title}{Effects of viscous dissipation during forced
  convection of power-law fluids in microchannels},
\newblock \bibinfo{journal}{International Communications in Heat and Mass
  Transfer} \bibinfo{volume}{89} (\bibinfo{year}{2017})
  \bibinfo{pages}{83--90}.
\bibitem[{Jing et~al.(2017)Jing, Pan, and Wang}]{jing2017joule}
\bibinfo{author}{D.~Jing}, \bibinfo{author}{Y.~Pan}, \bibinfo{author}{X.~Wang},
\newblock \bibinfo{title}{Joule heating, viscous dissipation and convective
  heat transfer of pressure-driven flow in a microchannel with surface
  charge-dependent slip},
\newblock \bibinfo{journal}{International Journal of Heat and Mass Transfer}
  \bibinfo{volume}{108} (\bibinfo{year}{2017}) \bibinfo{pages}{1305--1313}.
\bibitem[{Lu et~al.(2020)Lu, Xu, Gong, Duan, and Chai}]{lu2020effects}
\bibinfo{author}{H.~Lu}, \bibinfo{author}{M.~Xu}, \bibinfo{author}{L.~Gong},
  \bibinfo{author}{X.~Duan}, \bibinfo{author}{J.~C. Chai},
\newblock \bibinfo{title}{Effects of surface roughness in microchannel with
  passive heat transfer enhancement structures},
\newblock \bibinfo{journal}{International Journal of Heat and Mass Transfer}
  \bibinfo{volume}{148} (\bibinfo{year}{2020}) \bibinfo{pages}{119070}.
\bibitem[{Barletta(1996)}]{BARLETTA199615}
\bibinfo{author}{A.~Barletta},
\newblock \bibinfo{title}{Fully developed laminar forced convection in circular
  ducts for power-law fluids with viscous dissipation},
\newblock \bibinfo{journal}{International Journal of Heat and Mass Transfer}
  \bibinfo{volume}{40} (\bibinfo{year}{1996}) \bibinfo{pages}{15--26}.
\bibitem[{Wang and Wang(2007)}]{wang2007influence}
\bibinfo{author}{H.~Wang}, \bibinfo{author}{Y.~Wang},
\newblock \bibinfo{title}{Influence of three-dimensional wall roughness on the
  laminar flow in microtube},
\newblock \bibinfo{journal}{International Journal of Heat and Fluid Flow}
  \bibinfo{volume}{28} (\bibinfo{year}{2007}) \bibinfo{pages}{220--228}.
\bibitem[{Barletta et~al.(2025)Barletta, Celli, Sphaier, Brand{\~a}o, Lazzari,
  and Ghedini}]{barletta2025shape}
\bibinfo{author}{A.~Barletta}, \bibinfo{author}{M.~Celli},
  \bibinfo{author}{L.~A. Sphaier}, \bibinfo{author}{P.~V. Brand{\~a}o},
  \bibinfo{author}{S.~Lazzari}, \bibinfo{author}{E.~Ghedini},
\newblock \bibinfo{title}{Shape uncertainty analysis of laminar forced
  convection in a round microchannel with viscous dissipation},
\newblock \bibinfo{journal}{Applied Thermal Engineering} \bibinfo{volume}{265}
  (\bibinfo{year}{2025}) \bibinfo{pages}{125536}.
\bibitem[{Sphaier et~al.(2025)Sphaier, Barletta, Celli, Brand{\~a}o, and
  Ghedini}]{sphaier2025laminar}
\bibinfo{author}{L.~A. Sphaier}, \bibinfo{author}{A.~Barletta},
  \bibinfo{author}{M.~Celli}, \bibinfo{author}{P.~V. Brand{\~a}o},
  \bibinfo{author}{E.~Ghedini},
\newblock \bibinfo{title}{Laminar forced convection in circular microchannels
  with slip-flow: Analysis of randomly distributed roughness},
\newblock \bibinfo{journal}{International Communications in Heat and Mass
  Transfer} \bibinfo{volume}{162} (\bibinfo{year}{2025})
  \bibinfo{pages}{108615}.
\bibitem[{Shah and London(1978)}]{shah1978laminar}
\bibinfo{author}{R.~K. Shah}, \bibinfo{author}{A.~L. London},
\newblock \bibinfo{title}{{Laminar Flow Forced Convection in Ducts}},
\newblock in: \bibinfo{booktitle}{Advances in Heat Transfer},
  \bibinfo{publisher}{Academic Press}, \bibinfo{year}{1978}.
\bibitem[{Mat(2024)}]{Mathematica}
\bibinfo{title}{Wolfram {R}esearch{,} {I}nc., {M}athematica, {V}ersion 14.2},
  \bibinfo{howpublished}{{\tt https://www.wolfram.com/mathematica}},
  \bibinfo{year}{2024}. \bibinfo{note}{Champaign, IL}.
\bibitem[{Manglik and Bergles(1988)}]{manglik1988laminar}
\bibinfo{author}{R.~M. Manglik}, \bibinfo{author}{A.~E. Bergles},
\newblock \bibinfo{title}{Laminar flow heat transfer in a semi-circular tube
  with uniform wall temperature},
\newblock \bibinfo{journal}{International Journal of Heat and Mass Transfer}
  \bibinfo{volume}{31} (\bibinfo{year}{1988}) \bibinfo{pages}{625--636}.
\bibitem[{Languri and Hooman(2011)}]{LANGURI2011139}
\bibinfo{author}{E.~M. Languri}, \bibinfo{author}{K.~Hooman},
\newblock \bibinfo{title}{Slip flow forced convection in a microchannel with
  semi-circular cross-section},
\newblock \bibinfo{journal}{International Communications in Heat and Mass
  Transfer} \bibinfo{volume}{38} (\bibinfo{year}{2011})
  \bibinfo{pages}{139--143}.
\bibitem[{Alassar(2014)}]{alassar2014fully}
\bibinfo{author}{R.~S. Alassar},
\newblock \bibinfo{title}{Fully developed forced convection through
  semicircular ducts},
\newblock \bibinfo{journal}{Journal of Thermophysics and Heat Transfer}
  \bibinfo{volume}{28} (\bibinfo{year}{2014}) \bibinfo{pages}{560--565}.
\bibitem[{Mukherjee et~al.(2017)Mukherjee, Gupta, and
  Chhabra}]{mukherjee2017laminar}
\bibinfo{author}{S.~Mukherjee}, \bibinfo{author}{A.~K. Gupta},
  \bibinfo{author}{R.~P. Chhabra},
\newblock \bibinfo{title}{Laminar forced convection in power-law and {B}ingham
  plastic fluids in ducts of semi-circular and other cross-sections},
\newblock \bibinfo{journal}{International Journal of Heat and Mass Transfer}
  \bibinfo{volume}{104} (\bibinfo{year}{2017}) \bibinfo{pages}{112--141}.
\bibitem[{Kyritsi-Yiallourou and Georgiou(2018)}]{kyritsi2018newtonian}
\bibinfo{author}{S.~Kyritsi-Yiallourou}, \bibinfo{author}{G.~C. Georgiou},
\newblock \bibinfo{title}{Newtonian {P}oiseuille flow in ducts of
  annular-sector cross-sections with {N}avier slip},
\newblock \bibinfo{journal}{European Journal of Mechanics-B/Fluids}
  \bibinfo{volume}{72} (\bibinfo{year}{2018}) \bibinfo{pages}{87--102}.

\end{thebibliography}

\end{document}